\documentclass[a4paper]{article}

\usepackage{latexsym}
\usepackage{amsmath}

\newtheorem{theorem}{Theorem} 
\newtheorem{proposition}{Proposition} 
\newtheorem{corollary}{Corollary} 
\newtheorem{lemma}{Lemma}

\newcommand{\ansi}{\Lambda} 
 
\newcommand{\ccansi}{\Lambda^{\dc}}
\newcommand{\cansia}{\Lambda^{+}} 
\newcommand{\cansiaa}{\Lambda^{-}} 
 
\newcommand{\ReP}{\mathop{\rm Re}\nolimits}
\newcommand{\Res}[1]{#1_r} 
\newcommand{\Ims}[1]{#1_i} 
\newcommand{\CH}{\mathop{\rm C}\nolimits} 
\newcommand{\SH}{\mathop{\rm S}\nolimits} 
\newcommand{\conj}[1]{\overline{#1}}
\renewcommand{\P}{{\cal P}} 
\newcommand{\Q}{{\cal Q}} 
\newcommand{\R}{{\cal R}} 
\newcommand{\M}{{\cal M}} 
\newcommand{\N}{{\cal N}} 

\newcommand{\dC}{{\rm \kern.24em
            \vrule width.05em height1.4ex depth-.05ex
            \kern-.28em C}} 
\newcommand{\dc}{{\rm \kern.24em
            \vrule width.05em height1ex depth-.05ex
            \kern-.22em C}}
\newcommand{\dN}{{\rm I\kern-.20em N}}
\newcommand{\dR}{\rm I\kern-.17em R} 

\newcommand{\al}{\alpha}     \newcommand{\be}{\beta}

\newcommand{\la}{\lambda}    \newcommand{\ro}{\rho}       
\newcommand{\si}{\sigma}

\newcommand{\als}{\widetilde\alpha} 
\newcommand{\bes}{\widetilde\beta}

\newcommand{\A}{{\cal A}}

\newcommand{\decf}{F=\al U-\widetilde\al*U} 
\newcommand{\decg}{G=\be V-\widetilde\be*V} 
\newcommand{\ext}{\wedge}  
\newcommand{\medio}{\frac{1}{2}} 
 
\newcommand{\sign}{\mathop{\rm sign}\nolimits} 
\newcommand{\tr}{\mathop{tr}\nolimits} 

\newcommand{\bo}{\bullet} 

\newcommand{\alb}{\al_{\bo}} 
\newcommand{\alsb}{\als_{\bo}} 
\newcommand{\malb}{\frac{\al_{\bo}}{2}} 
\newcommand{\malsb}{\frac{\als_{\bo}}{2}}

\title{COMPOSITION OF LORENTZ TRANSFORMATIONS IN TERMS 
OF THEIR GENERATORS} 
\author{Bartolom\'e COLL 
\footnote{Syst\`{e}mes de r\'{e}f\'{e}rence spatio--temporels; 
Observatoire de Paris -- CNRS UMR 8630;
61,  Avenue de l'Observatoire;
75014 Paris; France}
 and\\ 
 Fernando SAN JOS\'{E}  MART\'{I}NEZ 
 \footnote{Depto de Matem\'atica Aplicada;
 E.T.S.I. Agr\'onomos;
 Universidad Polit\'ecnica de Madrid;
 Avd. de la Complutense s/n;
 28040 Madrid; Spain;\newline
 Tel.: 34 91 336 58 30; e--mail: sanjose@mat.etsia.upm.es}} 
\date{ }

\begin{document}

\maketitle 

\begin{abstract}
Two-forms in Minkowski space-time may be considered as generators 
of Lorentz transformations. Here, the
covariant and general expression for the composition law
(Baker-Campbell-Hausdorff formula) of two Lorentz
transformations in terms of their generators is obtained.
Every subalgebra of the Lorentz algebra of such generators, up to one, 
may be generated by a sole
pair of generators. When the subalgebra is known, the above BCH
formula for the two two-forms simplifies. Its simplified
expressions for all such subalgebras are also given.


\end{abstract}

\section{Introduction} 

In Minkowski space-time, global Lorentz transformations are used to relate
{\em inertial observers}. In a general space-time, local Lorentz
transformations \cite{local}, along time-like congruences of curves or
space-like families of hypersurfaces, are used to relate arbitrary frames
to {\em comoving observers} \cite{comoving} or to {\em synchronizations}
\cite{synchro}.

In most of the problems, the transformations involved belong to the
{\em proper orthocronous} Lorentz group (its connected component of
the identity), so that they are univocally given by the exponential
of the elements of the Lorentz algebra \cite{expgr}. Local Lorentz
transformations of the space-time may thus be given by exponentials of
the two-forms \cite{two-form} of the space-time.

This representation by two-forms of local Lorentz transformations has the
important advantage of involving exclusively its {\em intrinsic elements}
\cite{intrinsic}; these are the elements in which the corresponding two-forms
decompose \cite{sec-form}. Nevertheless,  this important advantage is 
obscured by the practical and formal difficulties that arise in the composition 
of transformations, where the corresponding two-forms are related by the 
Baker-Campbell-Hausdorff (BCH) formula. It is the absence of a simple and 
compact expression for the BCH formula that originates these difficulties. 
The main purpose of this paper is to obtain such an expression.

The simpler example of classical rotations illustrates clearly
the above situation in the three-dimensional euclidean case. Let us remember it.
The nine components of a rotation matrix $R$ may be related in more or less
complicate forms to differents relative parametrizations \cite{euler},
but the intrinsic elements of the matrix are the rotation axis ${\bf u}$
and the rotation angle $\alpha$. In the exponential domain, the  element
${\bf r}$ of the rotation algebra corresponding to $R$ decomposes in the
form
${\bf r} = \alpha *{\bf u}$ with $\alpha = |{\bf r}|.$
Here $*$ is the dual operator asociated to the 
euclidean metric $\delta$ so that,  ${\bf u}$ being a vector, 
$*{\bf u}$ is a  two-form. And $|{\bf r}|= ({\bf r},{\bf
r})^{\frac{1}{2}}$ is the {\em module} of ${\bf r}$, with 
$({\bf r},{\bf r})$ $=-\frac{1}{2}\tr{\bf r}^2$ and 
${\bf r}{\bf s}$ is the induced  product on two-forms
(in local coordinates 
$({\bf r}{\bf s})_{\mu\nu}={\bf r}_{\mu\tau}{\bf s}^\tau_\nu$). 
Thus, one has for $R$ the expression:
\begin{equation}
\label{expr}
R=\exp{\bf r}=\delta-\frac{\sin\alpha}{\alpha}\,{\bf r}+
        \frac{1-\cos\alpha}{\alpha^2}\,{\bf r}^2.
\end{equation}
And conversely, starting from $R$, one obtains
\begin{equation}
\label{logR}
{\bf r} = \log R = \frac{\arcsin\rho/2}{\rho}\left({}^tR-R\right)
\end{equation}
where $\rho$ is given by
$$
\rho \equiv \sqrt{(1+\tr R)(3 - \tr R)}
$$
and ${}^tR$ denotes the transposed of $R$. In other words, the rotation angle
$\alpha$ and the rotation axis ${\bf u}$ of a rotation matrix $R$ are intrinsically
given by
$$
\cos \alpha = \frac{1}{2}(\tr R - 1), \quad\quad\quad\quad {\bf u} = \frac{1}{\rho}*({}^tR-R).
$$ 

Suppose now that we have another rotation matrix $S$ corresponding to
the rotation vector $\beta{\bf v},$ that is to say, to the rotation algebra element 
${\bf s} = \beta*{\bf v}$. The element ${\bf t}$ 
corresponding to the composed rotation $T = RS$ is given by the BCH-formula 
\cite{bch}:
\begin{eqnarray}
\label{r.s}
\lefteqn{{\bf t} = {\bf r}\bullet{\bf s}}\nonumber\\
  & & \\
&=&\frac{2\sigma}{\sin \sigma}
\left\{
\frac{1}{\alpha}\sin\frac{\alpha}{2}\cos\frac{\beta}{2}\,{\bf r}+
\frac{1}{\beta}\cos\frac{\alpha}{2}\sin\frac{\beta}{2}\,{\bf s}+
\frac{1}{\alpha\beta}\sin\frac{\alpha}{2}
          \sin\frac{\beta}{2}\,[{\bf r},{\bf s}]
\right\},\nonumber
\end{eqnarray}
where $\sigma$ is given by
$$
\cos\sigma=
   \cos\frac{\alpha}{2}\cos\frac{\beta}{2}-
   \gamma\sin\frac{\alpha}{2}\sin\frac{\beta}{2},
$$ 
$\gamma$ being the cosinus of the rotation axes, $\gamma=$
$({\bf u},{\bf v}) =$ $\frac{1}{\alpha \beta}({\bf r},{\bf s}),$ and
$[{\bf r},{\bf s}]$ being the Lie bracket of the two-forms ${\bf r}$ and
${\bf s}$, $[{\bf r},{\bf s}] = $ ${\bf r}{\bf s} - {\bf s}{\bf r}$.

In other words, if ${\bf r}$ and ${\bf s}$ are two rotations corresponding to the rotation vectors 
$\alpha{\bf u}$ and $\beta{\bf v},$ the rotation angle $\theta$ and the rotation axis $\bf w$ of their 
composition ${\bf t},$ 
${\bf t} =$ $ {\bf r}\bullet{\bf s}=$ $\theta *{\bf w},$ are given by

\begin{eqnarray}
\label{tetaw}
\theta&=&2\sigma,\quad \nonumber\\
&&\\
{\bf w}&=&\frac{1}{\sin\sigma} 
\left\{ 
\sin\frac{\alpha}{2}\cos\frac{\beta}{2}\,{\bf u}+ 
\cos\frac{\alpha}{2}\sin\frac{\beta}{2}\,{\bf v}+ 
\sin\frac{\alpha}{2}\sin\frac{\beta}{2}\,{\bf u}\times {\bf v} 
\right\},\nonumber 
\end{eqnarray}  
where ${\bf u}\times {\bf v}$ is the vector product.
 
In Minkowski space-time, the analogous of expression (\ref{expr}), that is 
to say, the general and covariant explicit form  of  
local Lorentz  transformations as  exponential of two-forms,  has been given
in \cite{csj90}, although some partial results were already known 
\cite{otrosexp}. Nevertheless, the analogous of expression  
(\ref{logR}) for the two-forms as logarithms  of  local Lorentz
transformations seems not to have been considered but in
\cite{csj90}. 

Here we shall obtain the analogous of expressions (\ref{r.s}) and 
(\ref{tetaw}) 
for Minkowski space-time, that is to say, the general and covariant 
expression  of the Baker-Campbell-Hausdorf composition $\bullet$ 
of two two-forms as well as the relations between the intrinsic elements 
of this composition and those of its factor
two-forms \cite{relpar}. 

The BCH composition  of two two-forms belongs to the algebra generated
by them.  The unic proper subalgebra of the three-dimensional 
rotation algebra being one-dimensional
(rotations about same axis), one has for it $[{\bf r},{\bf s}] = 0$, and this
is the sole case in which expression (\ref{r.s}) simplifies. The situation
is drastrically different for the Lorentz algebra. From its thirteen proper
subalgebras, we have shown \cite{tofealgr} that
{\em twelve} of them may be generated by a pair
of two-forms, so that the BCH formula for the Lorentz algebra may be
simplified in many cases. We give here {\em all} these simplified
expressions. 

Our results are well adapted to theoretical considerations as
well as to practical computations. They may be applied in all situations
in which Lorentz transformations are implied, global ones in special 
relativity or local ones in both, special and general relativity. And this,  
not only for the above mentioned problems of adapted observers or 
synchronizations, but also in the study of special decompositions 
\cite{wyk}, Thomas precesion \cite{thomas}, general equations of helices 
\cite{synge}, motion of charged particles in particular 
electromagnetic fields \cite{honig}, or the generalization of the binomial 
theorem \cite{morales}.

These results may be also useful for heuristic researchs in other
fields. For example, in non linear electromagnetic theory. 
Physically, algebras are seen as first (tangent)
aproximations or weak (little) perturbations. This suggests a
guiding idea for the search of non linear electromagnetic
equations:  to consider that the first object to be
``nonlinearized" are not Maxwell equations for the electromagnetic
field, but {\it the electromagnetic field} itself. Being today described
by  a two-form (element of the Lorentz algebra), the ``finite" or
``strong"  description of the electromagnetic field would be given 
by a Lorentz field tensor, its exponential \cite{anotheraffair}. 

The paper is organized as follows. The computation of exponentials, 
logarithms and BCH compositions being easier in  complex spaces,  
Section \ref{sec:bchpre} is devoted to remember the real and complex elements that 
we shall need as well as to the obtaining of the exponential of complex two-forms. 
Section \ref{sec:bchbch} contains the general and covariant expression of 
the BCH composition of two two--forms as a linear combination of  
them, of their commutator and of their duals, with the coefficients depending on
the invariants of the pair of two--forms. In Section \ref{sec:bcha} the simplified 
expressions for each of the twelve proper subalgebras of the Lorentz algebra that 
may be generated by a pair of two-forms are obtained. In particular, half of these twelve 
subalgebras have the remarkable property that the eigenvalues of the BCH 
composition are the sum of the corresponding eigenvalues of the factor two--forms. 
The characterization of these subalgebras is given in this section.

\section{Preliminaries} 
\label{sec:bchpre}
We denote by $\ccansi$ the complexification of $\ansi,$
space of two--forms on Minkowski space $M_4$: $\ccansi$ is the complex linear
space associated to $\ansi\times\ansi$ by the complex structure
$J(F,G)=(-G,F)$ for $F,G\in\ansi.$ Thus, $\ccansi$ is a 
$\dC-$linear space of  complex dimension $6$ with two relevant 
$\dC-$linear subspaces 
\begin{eqnarray*} 
\cansia&=&\{f\in\ccansi|*f=if\}=\{F-i*F|F\in\ansi\}\\ 
\cansiaa&=&\{f\in\ccansi|*f=-if\}=\{F+i*F|F\in\ansi\},  
\end{eqnarray*} 
where $*$ is the dual operator associated to the Lorentzian metric of $M_4.$
It is verified that $\ccansi=\cansia\oplus\cansiaa$ and
$\overline{\cansia}=\cansiaa,$ so 
$$\dim_{\dc}\cansia=\dim_{\dc}\cansiaa=3.$$ 

As $\ansi$ has a structure of Lie algebra 
(with the commutator defined by $[F,G]=FG-GF,$ 
where the product is defined in local coordinates by 
$(FG)_{\mu\nu}=F_{\mu\tau}G^\tau_\nu$), 
$\cansia$ can be endowed with a $\dC-$Lie algebra structure by the 
commutator in $\ccansi$
\begin{equation} 
\label{pre:cext} 
[F+iG,H+iK]=[F,H]-[G,K]+i\left\{[F,K]+[G,H]\right\} 
\end{equation}  
then, $\ccansi$ is a $\dC-$Lie algebra and $\cansia$
is a $\dC-$Lie subalgebra of $\ccansi.$ 

Simarly, $\cansia$ is a $\dC-$metric linear space with the
$\dC-$scalar product in $\ccansi$
\begin{equation} 
\label{pre:epext} 
(F+iG,H+iK)=(F,H)-(G,K)+i\left\{(F,K)+(G,H)\right\},   
\end{equation}  
where $(\cdot,\cdot)$ is the induced scalar product in $\ansi$ 
given by $(F,G)=-(1/2)\tr(FG),$ $\tr$ being the trace operator. 
\par 
If $A\in\cansia,$ there is only one $F\in\ansi$ such
that $A=F-i*F.$ When $F$ is regular we have the decomposition 
$\decf$ with $\al>0,$ $\als\in\dR$ and $U$ a unitary two-form 
($(U,*U)=0$ and $(U,U)=-1$). Then $\pm\al,$ $\pm i\als$ are the
eigenvalues of $F,$ the pair $\{U,*U\}$ is called its 
{\em geometry}, and it is verified that $A=(\al-i\als)(U-i*U).$
When $F$ is null, its eigenvalues are $0,$ its {\em geometry} 
is $\{F,*F\}$ and it is verified that $A=F-i*F.$ As, $(U-i*U,U-i*U)=-2$ 
and $(F-i*F,F-i*F)=0$ for $F$ null, every non vanishing $A\in\cansia$ 
admits a unique decomposition of the form 
$$A=\la_A\,C_A$$ 
where $\la_A\in\dC,$ its real part $\ReP(\la_A)\geq 0$ 
and $C_A\in\cansia$ with $(C_A,C_A)\in\{-2,0\},$ 
$\la_A=1$ when $(C_A,C_A)=0.$ 
\par 

A non vanishing $A\in\cansia$ is called {\em regular} when
$(A,A)\not=0$ and null otherwise, $C_A$ is its {\em geometry} and
the complex number $a=\sqrt{-\medio(A,A)}$ 
its {\em invariant}. 
Let us note that only when $A$ is regular the number $\la_A$ 
of the above decomposition and its invariant $a$ coincide.\par 

Let $A,B\in\cansia$ with $A=F-i*F$ and $B=G-i*G$ for
$F,G\in\ansi.$ The {\em mixed invariants} of $F$ and $G$ are 
$$\rho=(F,*G)\qquad\hbox{and}\qquad\si=(F,G).$$ 
The complex number $k$ such that $-2k=(A,B)$ will be called
the {\em mixed invariant} of $A$ and $B.$ It is easily verified that
$k=-(\si-i\rho).$ For a pair of elements $A$ and $B$ of $\cansia$
the complex numbers $a$ (the invariant of $A$), $b$ (the invariant
of $B$) and $k$ (the mixed invariant of $A$ and $B$) are called
the {\em invariants of the pair $A,$ $B.$}\par  

The following expression gives the relation between the invariant
of $[A,B]$ and the invariants of the pair $A,$ $B$  
\begin{equation} 
([A,B],[A,B])=8(a^2b^2-k^2).\label{pre:ic}
\end{equation} 
This result is obtained by a straightforward computation taking into account 
Lemma 3 of \cite{csj95}.\par  

The space of tensors with two covariant indices can be endowed with
an associative  algebra structure with identity element as well
as a Lie algebra structure in the standard way; using linear 
extensions as in (\ref{pre:cext}) and (\ref{pre:epext}) the
complexification of that space can also be endowed with an  associative
algebra structure with identity element and with a
$\dC-$Lie algebra structure. The expression for the product is 
$$ 
(M+iN)(P+iQ)=MP-NQ+i(MQ+NP).  
$$
From the identity $FG-*G*F=-(F,*G)g,$ with $F$ and $G$ in $\ansi,$
it is obtained for $A$ and $B$ in $\cansia$  
\begin{equation} 
AB-*B*A=AB+BA=-(A,B) g; 
\label{pre:proepro} 
\end{equation}  
where the first equality is a consequence of the fact that $*A=iA$ for the
elements of $\cansia.$  

The following results are oriented to the obtaintion of the exponential of a complex 
two--form. Next lemma can be proven using (\ref{pre:proepro}) and induction over $n.$  

\begin{lemma} For any two--form $A$ of $\cansia$ with invariant $a,$  one has
$$A^{2n}=a^{2n} g\qquad\hbox{and}\qquad A^{2n+1}=a^{2n} A\qquad 
(n\in\dN). 
$$ 
\end{lemma} 

Let us define the complex functions $(z\in\dC)$
\begin{equation}
\label{exp:CyS}
\CH(z)=\cosh z\qquad\hbox{and}\qquad\SH(z)=\left\{ 
                               \begin{array}{ll} 
                 \displaystyle\frac{\sinh z}{z}&z\not=0\\[6pt]   
                                1&z=0.  
                               \end{array} 
                               \right.  
\end{equation}
These are entire complex functions. 

When $A\in\cansia$ it is verified that 
$$
\exp A=\sum_{n=0}^{\infty}\frac{A^n}{n!}
         =\sum_{n=0}^{\infty}\frac{A^{2n}}{2n!}+ 
           \sum_{n=0}^{\infty}\frac{A^{2n+1}}{(2n+1)!}
        =\cosh A+\sinh A; 
$$
last equality is a consequece of the definition of the hyperbolic
sine and cosine of a matrix. For these functions we have the
following result.  

\begin{proposition} 
For any two--form $A$ of $\cansia$ with invariant $a,$  one has
$$\cosh A=\CH(a) g\qquad\hbox{and}\qquad \sinh A=\SH(a) A.$$ 
\end{proposition} 

It can be proven as follows: 
\begin{eqnarray*} 
\cosh A&=&\sum_{n=0}^{\infty}\frac{A^{2n}}{2n!} 
       =\left(\sum_{n=0}^{\infty}\frac{a^{2n}}{2n!}\right)g  
       =\CH(a) g,\\  
\sinh A&=&\sum_{n=0}^{\infty}\frac{A^{2n+1}}{(2n+1)!}  
       =\left(\sum_{n=0}^{\infty}\frac{a^{2n}}{(2n+1)!}\right)A  
       =\SH(a) A. 
\end{eqnarray*} 

Then, as a corollary we obtain next theorem. 

\begin{theorem} 
For any two--form $A$ of $\cansia$ one has 
$$\exp A=\CH(a) g+\SH(a) A,$$ 
where $\CH$ and $\SH$ are the functions (\ref{exp:CyS}) of the invariant $a$ of $A$ and $g$ is the 
Lorentzian metric of the Minkowski space.
\end{theorem} 

Our expression of the exponential for the Lorentz group \cite{csj90} can be easely obtained from 
this result.

\section{The BCH--formula} 
\label{sec:bchbch}
The element $D\in\cansia$ such that $\exp A\exp B=\exp D$ or,
equivalently, $D=\log(\exp A\exp B)$ is given by the well known
BCH--formula; this defines the so called BCH composition 
$$A\bo B=\log(\exp A\exp B).$$ 
Then, we may write the complex version of our main result. 

\begin{theorem} 
\label{bch:bchc} 
The BCH composition $A\bo B$ of two two--forms $A,$ $B$ in $\in\cansia$ is given by
$$ 
A\bo B=\SH(d)^{-1}\left\{\SH(a)\CH(b) A+\CH(a)\SH(b) B+ 
	\frac{1}{2}\SH(a)\SH(b) [A,B]\right\}   
$$
where $d,$ the invariant of the two--form $A\bo B,$ is given by 
\begin{equation} 
\label{bch:invbch}
\cosh d=\CH(a)\CH(b)+k\SH(a)\SH(b)
\end{equation}  
$a,$ $b,$ $k$ being the invariants of the pair $A,$ $B.$ 
\end{theorem} 

\underline{Proof:} Let us put $D=A\bo B$ being $d$ its invariant
then, 
\begin{eqnarray*} 
\exp D&=&\CH(d) g+\SH(d) D=\exp A\exp B=(\CH(a) g+\SH(a) A)(\CH(b) g+\SH(b) B)\\ 
      &=&\CH(a)\CH(b)g+\SH(a)\CH(b)A+\CH(a)\SH(b)B+\SH(a)\SH(b)AB. 
\end{eqnarray*} 
First equality implies that we need to compute the antisymmetric
part of last expression to obtain $D$ as a function of $A$ and $B.$
If $Z$ stands for the antisymmetric part, we have 
$$Z=\SH(a)\CH(b)A+\CH(a)\SH(b)B+\medio\SH(a)\SH(b)[A,B],$$ 
and then 
$D=\SH(d)^{-1}\left\{\SH(a)\CH(b) A+\CH(a)\SH(b) B+ 
	\frac{1}{2}\SH(a)\SH(b) [A,B]\right\};$ 
so we have to obtain $d$ as a
function of the invariants of $A$ and $B.$ 
When $Z$ is regular, we have $\sinh d\, C_D=z C_Z,$  
for $C_Z$ and $C_D,$ the geometries of $Z$ and $D,$ respectively, and $z$ 
the invariant of $Z.$ So this equation yields $C_D=C_Z$ and 
$\sinh d= z.$ When $Z$ is null, $z=d=0$ and last expression is still true. 

Let us obtain the invariant $z$ as a function of $a,$ $b$ and $k.$ From
expressions (\ref{pre:ic}) and 
$$(A,[A,B])=(B,[A,B])=0,$$ 
derived from (10) of \cite{csj95}, we have 
\begin{eqnarray*} 
z^2&=&-\medio(Z,Z)\\ 
   &=&\SH(a)^2\CH(b)^2 a^2+\CH(a)^2\SH(b)^2 b^2 
	    +2\SH(a)\CH(b)\CH(a)\SH(b) k-\\
   & &\qquad \SH(a)^2\SH(b)^2(a^2b^2-k^2)\\ 
   &=&\left(\CH(a)\CH(b)+k\SH(a)\SH(b)\right)^2-1. 
\end{eqnarray*} 
Therefore,  we obtain the expression of the
theorem.$\Box$ 

To obtain the expression of the BCH--formula for the
Lorentz group we need the following proposition whose proof is
based on the fact that $[A,B]=0$ whenever $A\in\cansia$ and $B\in\cansiaa.$ 
 
\begin{proposition} 
\label{bch:re} 
Let $F,G\in\ansi$ and $A,B\in\cansia$ be such that 
$A=\medio(F-i*F)$ and $B=\medio(G-i*G),$ then 
$$F\bo G=2\ReP(A\bo B).$$ 
\end{proposition} 

\underline{Proof:} As, by definition 
$$\exp F\bo G=\exp F\exp G$$ 
then 
\begin{eqnarray*} 
\exp F\bo G&=&\exp(A+\conj{A})\exp(B+\conj{B})=\exp A\exp\conj{A}\exp B\exp\conj{B}\\ 
	   &=&\exp A\exp B\exp\conj{A}\exp\conj{B}=\exp A\exp B\,\conj{\exp A\exp B}\\ 
	   &=&\exp A\bo B\,\conj{\exp A\bo B}=\exp A\bo B\exp\conj{A\bo B}\\ 
	   &=&\exp\left((A\bo B)\bo\conj{(A\bo B)}\right)=\exp\left((A\bo B)+\conj{(A\bo B)}\right);  
\end{eqnarray*} 
second equality is due to the fact that they commute, last equality 
is because the only terms of 
the BCH--series of any elements $A$ and $B$ with $[A,B]=0$ 
is $A+B.$ $\Box$ 

With notation of the Theorem \ref{bch:bchc}, let us define the complex
functions 
\begin{eqnarray} 
\label{bch:bchrf} 
\P&=&\SH(d)^{-1}\SH(a)\CH(b)\nonumber\\ 
\Q&=&\SH(d)^{-1}\CH(a)\SH(b)\\ 
\R&=&\medio\SH(d)^{-1}\SH(a)\SH(b)\nonumber,  
\end{eqnarray}  
of the inavariants of the pair $A,$ $B$ and denote for short $X_r$ and $X_i$ the real and imaginary parts 
respectively of any complex function $X.$ A straightforward computation gives the main result for the real
case as a corollary of the previous theorem. 

\begin{theorem} 
\label{bch:er} 
The BCH composition $F\bo G$ of two real two--forms $F$ and $G$ on Minkowski space, is 
given by
$$
F\bo G=\Res{\P} F+\Res{\Q} G+\Res{R}[F,G]+\Ims{\P} *F+\Ims{\Q} *G+\Ims{R}*[F,G],  
$$
$\P,$ $\Q$ and $\R$ are the functions
(\ref{bch:bchrf}) of  the invariants of the pair
$(1/2)(F-i*F),$ $(1/2)(G-i*G).$ 
\end{theorem} 

The eigenvalues $\alb$ and $\alsb$ of $F\bo G,$ as well as the real and imaginary 
part of the functions $\P,$ $\Q$ and $\R$ may be expressed in terms of the real invariants of 
the pair of two--forms $F,$ $G.$ In that respect let us define the auxiliary functions of the 
invariants of the two--forms $F$ and $G,$ 
$\al,$ $\als,$ $\be$ and $\bes:$
\begin{eqnarray*} 
Cc^{\pm}&=&\cosh\frac{\al+\be}{2}\cos\frac{\als+\bes}{2}\pm 
          \cosh\frac{\al-\be}{2}\cos\frac{\als-\bes}{2}\\[5pt]      
Ss^{\pm}&=&\sinh\frac{\al+\be}{2}\sin\frac{\als+\bes}{2}\pm 
          \sinh\frac{\al-\be}{2}\sin\frac{\als-\bes}{2}\\[5pt]    
Cs^{\pm}&=&\cosh\frac{\al+\be}{2}\sin\frac{\als+\bes}{2}\pm 
           \cosh\frac{\al-\be}{2}\sin\frac{\als-\bes}{2}\\[5pt]    
Sc^{\pm}&=&\sinh\frac{\al+\be}{2}\cos\frac{\als+\bes}{2}\pm 
           \sinh\frac{\al-\be}{2}\cos\frac{\als-\bes}{2}  
\end{eqnarray*}  
and 
\begin{eqnarray*}
p&=&-\si(\al_m\be_m-\als_m\bes_m)-\ro(\al_m\bes_m+\als_m\be_m)\\[5pt]      
q&=&-\si(\al_m\bes_m+\als_m\be_m)+\ro(\al_m\be_m-\als_m\bes_m)   
\end{eqnarray*} 
where the {\it modular} scalars $\al_m$ y $\beta_m$ are given by 
$$
\al_m\equiv \frac{\al}{\al^2+\als^2},\qquad\be_m\equiv \frac{\be}{\al^2+\als^2}.
$$
Let us note that if $f$ is any of the functions $Cc^{\pm},$ $Ss^{\pm},$ $p$ or $q$ it is verified 
that 
\begin{equation} 
\label{bch:pp} 
f(\al,\als,\be,\bes)=f(\be,\bes,\al,\als); 
\end{equation} 
meanwhile, if $f^{\pm}$ is any of the functions $Cs^{\pm}$ or $Sc^{\pm},$ one has 
\begin{equation*}  
f^{\pm}(\al,\als,\be,\bes)=f^{\mp}(\be,\bes,\al,\als).
\end{equation*} 

Let us introduce the functions $\la,$ $\mu,$ $l$ and $m$ given by 
\begin{eqnarray}
&&\lambda=(1/2)(Cc^++pCc^-+qSs^-),\qquad \mu=(1/2)(Ss^++pSs^--qCc^-) \nonumber\\[10pt]
&&l=\lambda^2+\mu^2,\qquad m=\sqrt{(l+1)^2-4\lambda^2}. \label{bch:lm}
\end{eqnarray} 
From  Theorem \ref{bch:bchc}, we have 
\begin{equation*} 
\cosh\displaystyle\medio(\al_\bo-i\als_\bo)=\CH(a)\CH(b)+k\SH(a)\SH(b), 
\end{equation*} 
that is to say,  
$$
\lambda={\displaystyle\cosh\malb\cos\malsb},\qquad
\mu={\displaystyle\sinh\malb\sin\malsb}.
$$
Then, denoting by $\epsilon_\lambda$ and $\epsilon_\mu$ respectively 
the signs of the above scalars $\lambda$ and $\mu,$ one has
\begin{theorem} 
\label{bch:eigen}
The invariants $\alb$ and $\alsb$ of the BCH composition two--form $F\bo G$ 
of the two two--forms $F$ and $G,$ are given by 
$$
{\displaystyle\cosh\alb}=l+m,\qquad
{\displaystyle\cos\alsb}=l-m,
$$
where 
$\quad\sign(\sin\alsb)=\epsilon_\lambda\epsilon_\mu$ and, $l$ and $m$ are the 
functions (\ref{bch:lm}) of the invariants of the pair $F,$ $G.$
\end{theorem}

From this theorem and property (\ref{bch:pp}) one has the following result:
\begin{corollary}
The invariants of the BCH compositions $F\bo G$ and $G\bo F$ of two two--forms $F$ and 
$G$ conincide.
\end{corollary}

The real and imaginary part of $\SH(d)^{-1}$ are given, respectively, by 
\begin{eqnarray} 
\M &=&\frac{\displaystyle\alb\sinh\malb\cos\malsb+ 
                         \alsb\cosh\malb\sin\malsb} 
  {\displaystyle\sinh^2\malb\cos^2\malsb+ 
         \cosh^2\malb\sin^2\malsb}\nonumber\\
&&\label{bch:MN}\\
\N &=&\frac{\displaystyle\alb\cosh\malb\sin\malsb- 
                        \alsb\sinh\malb\cos\malsb} 
  {\displaystyle\sinh^2\malb\cos^2\malsb+\cosh^2\malb\sin^2\malsb}; \nonumber
\end{eqnarray} 
but from theorem \ref{bch:eigen} one has 

\begin{center}
\begin{tabular}{ll}
$\displaystyle\cosh(\malb)=(1/\sqrt{2})\sqrt{l+m+1},$&
                      $\displaystyle\cos(\malsb)=\epsilon_\la(1/\sqrt{2})\sqrt{l-m+1},$ \\[10pt]
$\displaystyle\sinh(\malb)=(1/\sqrt{2})\sqrt{l+m-1},$&
                       $\displaystyle\sin(\malsb)=\epsilon_\mu(1/\sqrt{2})\sqrt{-l-m+1},$
\end{tabular}
\end{center}
so that we have:

\begin{corollary}
The functions $\M$ and $\N$ of the invariants of the pair $F,$ $G$ are given by
\begin{eqnarray*}
\M&=&\frac{\epsilon_\la}{m}\sqrt{m+n}\;\arg\cosh(l+m)+\frac{\epsilon_\mu}{m}\sqrt{m-n}\;\arg\cos(l-m)\\
\N&=&\frac{\epsilon_\mu}{m}\sqrt{m-n}\;\arg\cos(l-m)+\frac{\epsilon_\la}{m}\sqrt{m+n}\;\arg\cosh(l+m).
\end{eqnarray*}
\end{corollary}

Then, one may obtain from (\ref{bch:bchrf}) the following result: 
\begin{theorem}
\label{bch:CoeficientesReales}
The coefficients of the BCH composition $F\bo G$ of two two--forms $F$ and $G$ in the 
expression 
$$
F\bo G=\Res{\P} F+\Res{\Q} G+\Res{R}[F,G]+\Ims{\P} *F+\Ims{\Q} *G+\Ims{R}*[F,G]  
$$
are given by

\begin{center} 
\begin{tabular}{rcl} 
&&\\[-6pt] 
$\Res{\P}$&$=$&$(\al_m \M-\als_m \N )Sc^++(\als_m \M +\al_m \N )Cs^+$\\[10pt]   
$\Ims{\P}$&$=$&$(\als_m \M +\al_m \N )Sc^+-(\al_m \M -\als_m \N )Cs^+$\\[10pt]  
$\Res{\R}$&$=$&$\left((\al_m\be_m-\als_m\bes_m)\M -(\al_m\bes_m+\als_m\be_m)\N\right)Cc^-+$\\[10pt]
  & &$\left((\al_m\bes_m+\als_m\be_m)\M +(\al_m\be_m-\als_m\bes_m)\N \right)Ss^-$\\[10pt]  
$\Ims{\R}$&$=$&$\left((\al_m\bes_m+\al_m\bes_m)\M +(\al_m\be_m-\als_m\bes_m)\N\right)Cc^- -$\\[10pt] 
  & &$\left((\al_m\be_m-\als_m\bes_m)\M -(\al_m\bes_m+\als_m\be_m)\N \right)Ss^-.$\\[10pt]  
$\Res{\Q}(\al,\als,\be,\bes)$&$=$&$\Res{\P}(\be,\bes,\al,\als)$\\[10pt]  
$\Ims{\Q}(\al,\als,\be,\bes)$&$=$&$\Ims{\P}(\be,\bes,\al,\als).$\\[10pt]  
\end{tabular}  
\end{center} 
\end{theorem} 

One could directly obtain the geometry of $F\bo G$ as a function of the invariants 
and the geometries of the pair $F,$ $G.$ In the singular case, the two--form and its geometry 
are proportional, so that we have only to study the regular case. Both cases are 
discriminated by the value of $\CH(a)\CH(b)+k\SH(a)\SH(b);$ from Theorem 
\ref{bch:bchc}, this value is $1$ only when $F\bo G$ is singular or 
$0.$ Taking into account Proposition \ref{bch:re}, that is to say,  
$$A\bo B=\medio\{F\bo G-i*(F\bo G)\},$$ 
and the definition of the geometry $\{W_{\bo},*W_{\bo}\}$ of $F\bo G,$
$$F\bo G=\al_\bo W_\bo-\als_\bo*W_\bo$$ 
one has
\begin{equation*}  
A\bo B=\medio(\al_\bo-i\als_\bo)(W_\bo-i*W_\bo); 
\end{equation*} 
then, from theorem \ref{bch:bchc} we obtain 
\begin{theorem} 
The element $W_{\bo}$ of the geometry $\{W_{\bo},*W_{\bo}\}$ of the BCH composition 
$F\bo G$ of two two-forms $F$ and $G,$ is given by the same expression of Theorem 
\ref{bch:CoeficientesReales} for $F\bo G$ where the functions (\ref{bch:MN}) $\M$ and 
$\N$ are substituted respectively by 
$$\medio\left(\M-\N\right)\quad\hbox{and}\quad\medio\left(\M+\N\right).$$
\end{theorem}

\section{Reduction of the BCH--formula for each type of subalgebra} 
\label{sec:bcha}
We have shown elsewhere \cite{tofealgr} that from the thirteen proper 
subalgebras of the Lorentz algebra, twelve of them may be generated by a pair of 
two-forms, so that the general expression of the BCH composition given above may 
be simplified in these twelve cases. It is the purpose of this Section to obtain the 
corresponding simplified expressions.

Half of these twelve cases have in common a particular property: {\em the 
eigenvalues $\alb,$ $\alsb$ of the BCH composition $F\bo G$ are the 
sum of the corresponding eigenvalues $\al,$ $\als$ of $F$ and $\be,$ $\bes$ of }$G:$
\begin{equation}
\label{bcha:SumaAutovalores}
	\alb = \al + \be,\qquad \alsb = \als + \bes.
\end{equation}

We shall begin by obtaining some common properties to these cases. From 
(\ref{bcha:SumaAutovalores}), expressions (\ref{bch:bchrf}) simplify to:
\begin{eqnarray} 
\label{bcha:bchrf+} 
\P&=&\frac{(a+b)\sinh a\cosh b}{a\sinh(a+b)}\nonumber\\ 
\Q&=&\frac{(a+b)\cosh a\sinh b}{b\sinh(a+b)}\\ 
\R&=&\medio\frac{(a+b)\sinh a\sinh b}{ab\sinh(a+b)}\nonumber.  
\end{eqnarray}  
So, we do not have to use (\ref{bch:invbch}) to obtain these coefficients. 
This expression implies that $d=a+b$ iff $k/ab=1.$ But 
$$
k=-\la_A\la_B(\si'-\ro'),
$$ 
where $\la_A$ is the invariant of $A$ when $A$ is regular and $1$ otherwise, 
$\si'=(U,V)$ and $\ro'=(U,*V)$ when $A=\medio(F-i*F)$ and $B=\medio(G-i*G).$ 
Therefore, 
$$\cosh d=
  \left\{ 
  \begin{array}{cl}
    \cosh a\cosh b -(\si'-i\ro')\sinh a\sinh b & \text{when both are regular}\\
    \cosh a        -(\si'-i\ro')\sinh a        & \text{when only $A$ is regular}\\
                   -(\si'-i\ro')               & \text{when both are null.} 
  \end{array} 
  \right. 
$$
Hence, $d=a+b$ iff $\ro'=0$ and $\si'=-1$ with $ab\not= 0,$ $\si'=0$ with 
$ab=0.$ These situations may be related to the relative positions of the non space 
like planes asociated to the geometries of the two--forms, that is, the pair $(\pi(U),\pi(V)).$ Relative positions of pairs of planes were studied in \cite{tofealgr}. With that notation, the above situations correspond to the following relative positions: 
$$
\Pi_{1,1}\;\Pi_{1,2}\;\Pi_{2,1},\;\Pi_{2,2},\;
\Pi_{\underline{2},2},\;\Pi_{\underline{2},3}.
$$ 
But these are the sole relatives positions of a pair of planes having at 
least one common null direction. Following the Schell classification of subalgebras of the 
Lorentz group \cite{sch61} (see Table \ref{tab:typesubalg} for the relation of
this notation and that of Patera {\em et al.} \cite{pwz75}),  our Theorem 3 of \cite{tofealgr} 
implies that a pair of two-forms, whose non spacelike planes asociated with their geometries 
have one of those positions, generates one and only one of the following types of subalgebra 
of dimension greater than two: 
$$R_6,\; R_7,\; R_8,\; R_{11},\; R_{12},\; R_{14}.$$ 
On the other hand, it may be verified that these are the sole algebras generated by two 
two--forms whose Killing--Cartan form is singular. Thus, denoting by $\A(F,G)$ the algebra 
generated by $F$ and $G,$ one has  
\begin{theorem} 
\label{bch:suma}
Let $\al,$ $\als$ be the eigenvalues of $F,$
$\be,$ $\bes$ those of $G,$ and $\alb,$ $\alsb$ those of their BCH composition $F\bo G,$ 
and let $\{U,*U\}$ and $\{V,*V\}$ be the geometries of $F$ and $G,$ respectively. Then, the 
following conditions are equivalent:
\begin{enumerate} 
\item $\al_\bo=\al+\be$ and $\als_\bo=\als+\bes,$ 
\item planes $\pi(U)$ and $\pi(V)$ have at least one common null direction,
\item the relative position of the pair of plane $(\pi(U),\pi(V))$ is one and only one of the 
folowing: $\Pi_{1,1}\;\Pi_{1,2}\;\Pi_{2,1},\;\Pi_{2,2},\;
\Pi_{\underline{2},2},\;\Pi_{\underline{2},3},$ 
\item the algebra $\A(F,G)$ is one of the following ones:
\newline 
$R_6,\; R_7,\; R_8,\; R_{11},\; R_{12},\; R_{14},$ 
\item the Killing--Cartan form of $\A(F,G)$ is singular.
\end{enumerate} 
\end{theorem} 

\begin{table} 
\centering
\begin{tabular}{|c|l|l|l|} 
\hline
Dimension&\multicolumn{2}{c|}{Type}&\multicolumn{1}{c|}{Basis}\\
\hline
$0$&$R_1$&$F_{15}$& \\ \hline 
$1$&$R_2$&$F_{13}$&$A_1$\\ 
$1$&$R_3$&$F_{14}$&$A_3$\\ 
$1$&$R_4$&$F_{12}$&$A_2$\\ 
$1$&$R_5$&$F_{11}$&$A_1+\tau A_2\ (\tau\not=0)$\\ \hline 
$2$&$R_6$&$F_{8}$&$A_1,A_3$\\ 
$2$&$R_7$&$F_{9}$&$A_1,A_2$\\ 
$2$&$R_8$&$F_{10}$&$A_3,A_4$\\ \hline 
$3$&$R_{9}$&$F_{7}$&$A_1,A_3,A_4$\\ 
$3$&$R_{10}$&$F_{4}$&$A_1,A_3,A_5$\\ 
$3$&$R_{11}$&$F_{6}$&$A_2,A_3,A_4$\\ 
$3$&$R_{12}$&$F_{5}$&$A_1+\tau A_2,A_3,A_4\ (\tau\not=0)$\\ 
$3$&$R_{13}$&$F_{3}$&$A_2,A_3-A_5,A_4-A_6$\\ \hline 
$5$&$R_{14}$&$F_{2}$&$A_1,A_2,A_3,A_4$\\ \hline 
$6$&$R_{15}$&$F_{1}$&$A_1,A_2,A_3,A_4,A_5,A_6$\\ \hline 
\end{tabular}
\caption{Subalgebras of the Lie algebra of Lorentz group}
\label{tab:typesubalg}
\end{table}

The subalgebras of the Lie algebra of Lorentz group are given in Table 
\ref{tab:typesubalg}. Each type  corresponds to a
class of conjugation by the orthochronous proper Lorentz  
group (connected component of the identity),
except  for the types where it appears the real number 
$\tau.$ In these cases each $\tau$ defines a different conjugation class. 
The $R$'s and the $F$'s columns of Table \ref{tab:typesubalg} correspond, 
respectively, to notations by Schell (see \cite{sch61}) and by Patera et al. 
(see \cite{pwz75}). The last column includes a basis of each type in 
terms of the two--forms associated to a real null tetrad 
$\{l,m,p,q\}$ (that is to say a tetrad verifying
$(l,l)=(m,m)=(l,p)=(l,q)=(m,p)=(m,q)=(p,q)=0$ and
$(l,m)=-(p,p)=-(q,q)=1$);
we have made use of the notation: 
\begin{equation*} 
\begin{array}{ccc} 
A_1=l\ext m&A_3=l\ext p&A_5=m\ext p\\ 
A_2=p\ext q&A_4=l\ext q&A_6=m\ext q.  
\end{array} 
\end{equation*} 

From now on it is considered that 
$$A=\medio(F-i*F)\qquad\hbox{and}\qquad B=\medio(G-i*G).$$ 

When $\A(F,G)$ is a $R_7$ or $R_8$ algebra 
then $[F,G]=0$ and $[A,B]=0.$ In these cases $d=a+b$ (Theorem \ref{bch:suma}). 
For $R_8$ both are null and then $\P=\Q=\R=1;$ therefore, $F\bo G=F+G.$ 
For $R_7$ both are regular and the geometries of $A$ and $B$ can be taken such that
$C_A=C_B$ (Proposition 8 of \cite{csj95}); hence $bA=aB.$
Therefore, 
$$
A\bo B=\frac{a+b}{\sinh(a+b)} \left\{\frac{\sinh a}{a}\cosh{b}\, A+
      \cosh{a}\frac{\sinh b}{b}\,B\right\}=\left(1+\frac{b}{a}\right) A=A+B.
$$

Thus, the well--known result for commuting two-forms is obtained; 

\begin{theorem} When $\A(F,G)$ is a $R_7$ or $R_8$ algebra 
one has $F\bo G=F+G.$ 
\end{theorem} 

Suppose $\A(F,G)$ is a $R_6$ algebra then 
$\dim\A(F,G)=2$ and $[F,G]\not=0.$ Thus, from Theorem 3 of
\cite{tofealgr}, $F$ and $G$ are simple two-forms, that have only
one common  principal direction and at least one non common
principal one. So, at least one is regular (say $F$), $a,$ $b$ and $k$ are real 
and, by Theorem \ref{bch:suma}, $d=a+b.$ If $\al$ is the non null eigenvalue of $F$ 
a straightforward computation yields 
\begin{equation} 
\label{a:a6com} 
[F,G]=\frac{\si}{\al}F+\al G.
\end{equation} 
Let us remark, that $\si=-1$ when both two--forms are regular and $\si=0$ when one is 
null ($G$ in this case). 
Taking into account expressions (\ref{bcha:bchrf+}) and (\ref{a:a6com})
the following theorem is obtained, 

\begin{theorem} When $\A(F,G)$ is a $R_6$ algebra and $F$ is the regular two--form 
one has 
$$
   F\bo G=\frac{\al+\be}{\sinh\frac{\al+\be}{2}}
   \left\{
   \frac{\sinh\frac{\al}{2}}{\al}\left(\cosh\frac{\be}{2}-\sinh\frac{\be}{2}\right)+
   \frac{\sinh\frac{\be}{2}}{\be}\left(\cosh\frac{\al}{2}+\sinh\frac{\al}{2}\right)G 
   \right\}
$$ 
where $\al$ is the non null eigenvalue of $F$ and $\be$ the eigenvalue of $G.$ 
\end{theorem} 

For $R_{10},$ $R_{11}$ and $R_{13}$ algebras we 
know from Theorem 3 of \cite{tofealgr} that $F$ and $G$ are simple 
and $(F,*G)=0.$ First condition implies that $a$ and $b$ are 
pure real or imaginary numbers; then functions $\CH(a),$ $\SH(a),$ 
$\CH(b)$ and $\SH(b)$ are real valued. Second 
condition implies that $k$ is real. Moreover, $d/\sinh d$ is real. 
Let us show this. As $F\bo G$ is a linear combination of $F,$ $G$ and 
$[F,G],$ $(F\bo G,*F\bo G)=0$ because $[F,G]$ is simple (Proposition 9 
of \cite{csj95}) and $[F,[F,G]]=[F,*[F,G]]=0;$ thus $d$ is a pure real or 
imaginary number which implies that $d/\sinh d$ is real. Therefore, $\P,$ 
$\Q$ and $\R$ are real valued functions of complex variables. 

\begin{theorem} 
When $\A(F,G)$ is a $R_{10},$ $R_{11}$ or $R_{13}$ algebra, one has  
$$F\bo G=\P F+\Q G+\R[F,G].$$ 
\end{theorem}
 
Let us remark that this expression is equivalent to (\ref{r.s}) for
$R_{13},$ this algebra being ismorphic to the Lie algebra of the
rotation group in $\dR^3.$\par 

Consider $\A(F,G)$ is a $\R_{12}$ algebra. 
In this case (Theorem 3 of \cite{tofealgr}) $F$ and $G$ are not both simple,
they have one common principal direction and, either  their eigenvalues ar proportional  or 
one is null and the other is not simple. Assume $F$ is not simple. Therefore, 
$d=a+b$ (Theorem \ref{bch:suma}) and there is a real number $\eta$ such that 
$b=\eta a$ ($\eta=0$ when G is null). Taking into account (\ref{bcha:bchrf+}), one has 
\begin{eqnarray*}  
\P&=&\frac{(1+\eta)\sinh a\cosh\eta a}{\sinh(1+\eta)a}\nonumber\\ 
\Q&=&\frac{(1+\eta)\cosh a\sinh\eta a}{\eta\sinh(1+\eta)a}\\ 
\R&=&\medio\frac{(1+\eta)\sinh a\sinh\eta a}{\eta a\sinh(1+\eta)a}\nonumber.  
\end{eqnarray*}
Thus $\P+\eta\Q=1+\eta$ 
which implies 
\begin{equation} 
\label{a:ca12} 
\Res{\P}=1+\eta-\eta\Res{\Q}, \qquad\Ims{\P}=-\eta\Ims{\Q}.
\end{equation} 

For the non spacelike part of the geometries of $F$ and $G,$ $U$ and $V$ 
one obtain, from (\ref{a:a6com}), that $[U,V]=\si'U+V$ where $\si'=(U,V).$ 
As $b=\eta a,$ the commutator of $F$ and $G$ verifies 
\begin{equation} 
\label{a:com12}
[F,G]=-\al(\eta F-G)+\als*(\eta F-G).
\end{equation} 

Combining (\ref{a:ca12}) and (\ref{a:com12}) 
we obtain the following result. 

\begin{theorem} Let $\A(F,G)$ be a $\R_{12}$ algebra, and  
$\eta$ the real number such that $\be-i\bes=\eta(\al-i\als)$ for the eigenvalues 
of $G$ and $F;$ when $F$ is not simple one has 
\begin{eqnarray*} 
F\bo G&=&\left\{ 
 1-\eta
 \left(\Res{\Q}+\frac{\al}{\als}\,\Ims{\Q}+ 
       \frac{\al^2+\als^2}{\als}\,\Ims{\R}-1\right)  
         \right\} F+\\  
      & &\left\{\Res{\Q}+\frac{\al}{\als}\,\Ims{\Q} +
         \frac{\al^2+\als^2}{\als}\,\Ims{\R}\right\} G-\left\{\frac{1}{\als}\,\Ims{\Q}-\Res{\R}+
         \frac{\al}{\als}\,\Ims{\R}\right\}[F,G].  
\end{eqnarray*}        
\end{theorem}
 
When $\A(F,G)$ is a $R_{14}$ algebra, expression (\ref{a:a6com})
still holds for the non spacelike part of the geometries of $F$ and $G$; 
now it is verified that $\si_{UV}=-1;$ thus 
$$[F,G]=-\be F+\al G+\bes*F-\als*G.$$ 
Therefore, we have the following 
result. 

\begin{theorem} 
Let $\decf$ and $\decg$ be such that $\A(F,G)$ is a $R_{14}$ 
algebra then one has 
\begin{eqnarray*} 
F\bo G&=&(\Res{\P}-\be\Res{\R}-\bes\Ims{\R})F+(\Res{\Q}+\al\Res{\R}+\als\Ims{\R})G+\\ 
      & &(\Ims{\P}+\bes\Res{\R}-\be\Ims{\R})*F+(\Ims{\Q}-\als\Res{\R}+\al\Ims{\R})*G. 
\end{eqnarray*} 
\end{theorem} 

\section{Acknowledgement} 
F. San Jos\'e Mart\'{\i}nez is partially supported by DGICYT, 
PB92-0220; he would like to thank Prof. E. Aguirre Dab\'an  
for his interest.


\begin{thebibliography}{99} 
 
\bibitem{local} Here a {\em local Lorentz transformation} means a 
field of Lorentz transformations of the tangent space at {\em each} 
point. 

\bibitem{comoving} That is to say, to observers whose velocity  
vector is tangent to the curves of the congruence. 

\bibitem{synchro} A {\em synchronization} is a specification of 
the locus of points (hypersurfaces) of equal time. Every  
synchronization has a  {\em natural} set of observers: those whose 
velocity vector is normal to the family of hypersurfaces. 

\bibitem{expgr} They are given biunivocally only for  
{\em exponential} groups (J. Dixmier, Bull. Soc. math. France,  
{\bf 85}, 113 (1957); L. Pukanszky, Trans. Amer. Math. Soc., 
{\bf 126}, 487 (1967). 

\bibitem{two-form} Because of the space-time metric, the elements  
of the algebra may be written as  second order antisymmetric 
covariant tensors at each point so that, in the corresponding 
domain of the space-time, they define a two-form. 

\bibitem{intrinsic} The usual representation of Lorentz  
transformations by matrices or second order tensors carries an 
excessive number of nonstrict quantities, namely $ n^2=16 $ 
components, which hides their {\em intrinsic elements}. These 
elements  depend only on  
$ n(n-1)/2 = 6 $ parameters, the group dimension, and 
are those in which the Lorentz transformations may be  
{\em biunivocally} and {\em covariantly} decomposed. The
intrinsic    elements of a Lorentz transformation are thus its non 
space-like  invariant 2-plane, and its two eigenvalues. As it is 
well known,  in the regular case one of these eigenvalues is a 
hyperbolic angle, and gives the  magnitude of the {\em proper} 
boost on the timelike invariant 2-plane.   The other eigenvalue is 
a trigonometric angle, and fixes the rotation on the  orthogonal 
space-like invariant  2-plane. Intrinsic elements have not to be 
confused with velocity-rotation relative parametrizations, for 
which  biunivocity fails. 


\bibitem{sec-form} See next Section. 

\bibitem{euler} Euler angles, Cayley-Klein parameters, etc. 


\bibitem{bch} For the original articles see: J.E. Campbell, Proc. 
London Math. Soc., {\bf 28}, 381 (1897); {\bf 29}, 14 (1898);
H.  F. Baker,  {\em ibid}, {\bf 34}, 347 (1902); {\bf 2}, 293 (1904); 
{\bf 3}, 24 (1905); F. Hausdorff, Ber. Verhandl. 
Saechs. Akad. Wiss. Leipzig, Math. Naturw. Kl., {\bf 58}, 19 (1906).

\bibitem{csj90} B. Coll and F. San Jos\'e, Gen. Relativ. Gravit.,  
{\bf 22}, 811 (1990).

\bibitem{otrosexp} A. H. Taub, Phys. Rev., {\bf 73}, 786 (1948); 
S. L. Bazanski, J. Math. Phys., {\bf 6}, 1201 (1965); 

\bibitem{relpar} It is to be noted that the product of two Lorentz 
transformations in terms of {\em relative} parameters (namely, 
relative velocity and relative rotation), which is well known from 
long time (see, for exemple M. Rivas {\em et al.}, Eur. J. Phys.,  
{\bf 7}, 1 (1986)), differs strongly from the BCH product. This is 
due to the facts that in the case of relative parametrizations 
every factor is framed in a different basis (i.e. for different 
observers) and that the corresponding parametrizations 
refers to these different bases. A connection between the 
relative   product and the BCH formula not only needs the relation 
between relative parameters and intrinsic elements, but {\em also} 
the relation between the relative parameters with respect to 
different observers, involving notions such as ``velocity of a 
point with respect to an observer {\em as seen} by another 
observer". We shall not consider here such a connection. 

\bibitem{tofealgr} B. Coll and F. San Jos\'e, {\em On the 
algebras generated by two 2-forms in Minkowski space-time}, 
J. Math. Phys., {\bf 37}, 5792 (1996); see also {\em Relative 
position of a pair of planes and algebras generated by two  2-forms in 
relativity}, in {\em Recent Developments in Gravitation}, World 
Scientific, 1991, p. 210. 

\bibitem{wyk} See for example C. B. van Wyk, J. Math. Phys.,  
{\bf 32}, 425 (1991). 

\bibitem{thomas} Corrections to the Tomas precesion (L. H. 
Thomas, {\em Nature}, {\bf 117}, 514 (1926)) have been obtained 
from the BCH formula  by N. Salingaros, J. Math. Phys., {\bf 25}, 
706 (1984).

\bibitem{synge} The equations of relativistic  helices are due to 
J. L. Synge, Proc. Roy. Irish Acad., sec. A, {\bf 65}, 27 (1967); 
their expressions may by manifestly simplified using the results 
presented here. 

\bibitem{honig} The first to consider this problem in general 
was A. H. Taub, Phys. Rev., {\bf 73}, 786 (1948); and an intrinsic 
characterization in term of Frenet-Serret parameters was given by 
E. Honig {\em et al.}, J. Math. Phys., {\bf 15}, 774 (1974). 

\bibitem{morales} See J. Morales and A. Flores-Riveros, J. Math. 
Phys., {\bf 30}, 393 (1989). 

\bibitem{anotheraffair} The analysis of this idea needs, 
beside our  results, their reciprocals: the internal operations on the
group of Lorentz tensors that correspond by the exponential to the
two internal operations of its algebra  (addition and commutator). But
this is another affair. 





\bibitem{csj95} B. Coll, F. San Jos\'e  Mart\'{\i}nez; J. Math.
Phy., vol. {\bf 36}, 4350 (1995). 

\bibitem{sch61} J.F. Schell; J. Math. Phys., {\bf 2}, 202 (1961).

\bibitem{pwz75} J. Patera, P. Winternitz, H. Zassenhaus; J. Math. 
Phys., {\bf 16}, 1597 (1975). 


\end{thebibliography}
\end{document}